# Title: Soil and soil $CO_2$ magnify greenhouse effect

**Authors:** Weixin Zhang[1†], Chengde Yu[2†], Zhifeng Shen[1], Shu Liu[1], Suli Li[1], Yuanhu Shao[1], Shenglei Fu[1*].

**Affiliations:**

[1]Key Laboratory of Geospatial Technology for the Middle and Lower Yellow River Regions, Ministry of Education, College of Environment and Planning, Henan University, Kaifeng 475004, China.

[2]School of Life Sciences, Henan University, Kaifeng 475004, China.

[*]Correspondence to: sfu@scbg.ac.cn

[†]These authors contributed equally to this work.

## ABSTRACT

Soil has been recognized as an indirect driver of global warming by regulating atmospheric greenhouse gases. However, in view of the higher heat capacity and $CO_2$ concentration in soil than those in atmosphere, the direct contributions of soil to greenhouse effect may be non-ignorable. Through field manipulation of $CO_2$ concentration both in soil and atmosphere, we demonstrated that the soil-retained heat and its slow transmission process within soil may cause slower heat leaking from the earth. Furthermore, soil air temperature was non-linearly affected by soil $CO_2$ concentration with the highest value under 7500 ppm $CO_2$. This study indicates that the soil and soil $CO_2$, together with atmospheric $CO_2$, play indispensable roles in fueling the greenhouse effect. We proposed that anthropogenic changes in soils should be focused in understanding drivers of the globe warming.

**INTRODUCTION**

$CO_2$ is one of the major greenhouse gases and the change in atmospheric $CO_2$ concentration has been considered as the most important driver of global warming [1-3]. The low heat capacity of atmospheric air and the strong convection processes in the troposphere make it difficult to effectively retain heat energy within atmospheric air. The magnitude of greenhouse effect would be much limited if the atmospheric $CO_2$-absorbed surface radiation could not be largely retained within the earth. Previously, the role of soil (including soil water and gases) in the greenhouse effect has been recognized as sources and/or sinks of atmospheric greenhouse gases [4-8]. However, is it possible that soil and $CO_2$ in soil directly contribute to the greenhouse effect? There are several characteristics of soil that may facilitate soil to play essential roles in global warming. Firstly, surface soil (e.g., 0-20 cm) was likely to access heat energy either from solar radiation or from the surface downward longwave radiation. In view of the higher heat capacity and slower heat loss in soil than those in atmosphere [9], soil may act as one of the more efficient heat storages. The observed 31% greater increase of soil surface temperature than increase of air temperature in China (1962-2011) [10] may imply that the heat is more readily retained in soil than in atmosphere. The energy flux from soil can affect the land-atmosphere interactions and potentially regulate weather and climate [9,11]. Secondly, although the total volume of $CO_2$ in soil air is less than the volume of atmospheric $CO_2$, the concentrations of $CO_2$ in soil were usually 5-100 fold higher than that of the concentration of atmospheric $CO_2$ [12]. This highly-enriched $CO_2$ in soil may further alter the heat balance in surface lands. However, no studies have tried to evaluate the potential direct contributions of soil and soil $CO_2$ to greenhouse effect. We considered that the greenhouse effect would rely on both the roles of atmospheric $CO_2$ in absorbing heat radiation and the roles of soil and soil $CO_2$ in retaining heat. We hypothesized that an increase of $CO_2$ concentration in soils would enhance the process of heat trapping within soils, and, thus, regulate the greenhouse effect on the earth.

Here, we performed a field mesocosm study in which $CO_2$ concentration was manipulated at five levels (i.e., L300 ppm, L480 ppm, L3200 ppm, L7500 ppm and L16900 ppm) and air temperature in mesocosms that standing in atmosphere (simulated $AT_{atm}$) or being buried under 10 cm of surface soil layer (simulated $AT_{soil}$) were monitored for six days.

**RESULTS**

**Temporal change pattern of air temperatures in atmosphere and soil**

To show the potential roles of atmospheric air and soil in heat balance, the daily amplitudes and temporal change pattern of air temperatures both in atmosphere and soil were examined. The simulated $AT_{atm}$ fluctuated from 15.95℃ to -3.70℃ during the six experimental days (Fig. 1). For a given day, the changes of simulated $AT_{atm}$ can be separated into four periods (Atom-P1 to Atom-P4) (see Methods; Fig. 1); the simulated $AT_{atm}$ started to increase during 9:11 – 9:51, occasionally at 11:21 (day 6), and reach the highest temperature during 12:31 – 13:21. While, the simulated $AT_{soil}$ only fluctuated from 5.42℃ to 1.22℃ (Fig. 2). For a given day, the changes of simulated $AT_{soil}$ can also be separated into four periods (Soil-P1 to Soil-P4) (see Methods; Fig. 2); but the simulated $AT_{soil}$ started to increase during 11:51 – 15:21, and reach the highest temperature during 18:41 – 21:01.

The temporal change pattern of temperature in the real atmospheric air was as similar as that in the simulated atmospheric air with $CO_2$ concentration of L300 but lower than that with $CO_2$ concentration of L480 during periods with higher heat radiation (Atom-P2 and Atom-P3); in contrast, temperatures in the real atmospheric air were consistently higher than those in both the two types of simulated atmospheric air during periods with lower heat radiation (Atom-P1 and Atom-P4) (Fig. S1). In general, the temperatures of simulated soil air could be as high as those of bulk soil during the short warmest periods in soil in the day, furthermore, the temperatures of simulated soil air with $CO_2$ concentration of L7500 could be higher than those of bulk soil on days when the total heat radiation was much higher; in addition, the temperatures of simulated soil air fluctuated with larger amplitudes and mostly were lower than those of the bulk soil (Fig. S2).

**$CO_2$ concentration effects on air temperatures in atmosphere and soil**

The effects of $CO_2$ concentration on air temperatures both in the atmospheric air and soil air were examined. In general, the simulated $AT_{atm}$ increased with $CO_2$ concentration significantly at periods of Atom-P2 and Atom-P3 especially during day 1 – day 5 when the sun-derived radiation was relatively higher (Table S1). A decline of the simulated $AT_{atm}$ in response to the increase of $CO_2$ concentration at periods of Atom-P2 and/or Atom-P3 was observed when compared with the temperatures in the treatment with L7500 to that with L16900 during day 1- day 3 (Table S1). Furthermore, a decline of the simulated $AT_{atm}$ in response to the increase of $CO_2$ concentration at periods of Atom-P1 and Atom-P4 occurred when comparing the temperature in the treatment with low $CO_2$ concentration (i.e., L300 or L480) to that with high concentration (i.e., L3200 or L7500 or L19600) (Table S1; Fig. S3). In brief, the simulated $AT_{atm}$ at periods of Atom-P1 and Atom-P4 was the highest in treatment with L480, which was significantly higher than that in treatment with L7500 and/or L19600.

The influences of soil $CO_2$ concentration on air temperature in soil were more complex than those in atmosphere. In general, the simulated $AT_{soil}$ increased with $CO_2$ concentration significantly at periods of Soil-P2 and/or Soil-P3 during day 1 – day 5 when the sun-derived radiation was relatively higher (Table S2). A decline of the simulated $AT_{soil}$ in response to the increase of $CO_2$ concentration at periods of Soil-P2 and/or Soil-P3 was also observed when compared the temperature in treatment with L7500 to that with L16900 during day 1- day 5 (Table S2). In brief, the simulated $AT_{soil}$ in treatment with L7500 was the highest, which was usually significantly higher than that with L480 and/or with L16900 at periods of Soil-P2 and/or Soil-P3 during day 1-day 5. In addition, the simulated $AT_{soil}$ in treatment with L7500 was only significantly higher than that with L3200 for stage of Soil-P2 at day 1, but was significantly higher than that in all other treatments of $CO_2$ concentration for stage of Soil-P2 at Day 3. On the contrary, we did not observe significant decline of the simulated $AT_{soil}$ in response to the increase of $CO_2$ concentration at periods of Soil-P1 and Soil-P4 when compared the temperature in the treatment with low $CO_2$ concentration (i.e., L300 or L480) to that with high concentration (i.e., L3200 or L7500 or L19600).

## DISCUSSION

**The complimentary roles of atmospheric air and soil in greenhouse effect**

In our study, the temperatures of both the real and simulated atmospheric air exerted great amplitudes indicating that the heat in the atmosphere was readily lost into outer space of the earth. We considered that the atmospheric $CO_2$ plays a role as "racket" that catches the "ball of surface radiation" and beats it out in all directions. Thus, a certain proportion of heat would escape to outer space for each round of $CO_2$-based radiation absorption and re-emission. In other words, if only the role of atmospheric $CO_2$ was considered, for example, when the surface lands were totally covered by bare rocks, the heat amount that the earth could retain would be much limited (Fig. 3A).

In contrast, one of the most distinct characteristics of soil is the slower heat transmission compared to that in atmosphere [13]. The temperatures of both the bulk soil and simulated soil air exerted smaller amplitudes than those in atmosphere indicating that the soil may possess higher heat capacity and hold heat more efficiently than atmospheric air. For instance, a simulation study using desert soils reported that 35% of the net radiation may be transferred into soil [14]; in addition, soil heat flux was found to result in 7.6% of the net radiation being stored in soil at daytime, and acted as a heat source to outer soil layers at night-time accounting for more than 50% of the net night-time radiation at an Antarctic area during warmer months [15]. Also, soil temperature was found to be generally higher than surface air temperature in the Tibetan Plateau during 1983-2013 [16]. Therefore, either the solar radiation or the "ball of surface radiation" that was kicked back to the surface soil by the atmospheric $CO_2$ could be partly retained in soil and thus potentially form a cache of heat (Fig. 3B). This heat storage in soil would contribute to the greenhouse effect on the earth.

Furthermore, it was notable that in this study soil air temperature reached the highest values $403 \pm 24$ min later than that of atmospheric air. The complex heat transfer either through conduction or convection among the soil solid phase, liquid water and soil gases [9] may cause such lagging warming in soil. Importantly, it retained heat within soil for a longer period of time. A recent study suggested that the cooling in the deep Pacific may offset more than one-fourth of the heat gain above 2000 m [17]. Here, however, the lagging warming of soil would reduce the possibility of heat loss into outer space during a given period and potentially provide an opportunity of transferring heat from soil to surface air during colder periods (Fig. 3B). Hence, other than ground radiation, soil heat flux would be an essential heat source to surface atmospheric air especially at night-time and early morning, which may increase the level of the daily minimum temperature in surface atmosphere.

Therefore, we considered that the greenhouse effect resulted from three closely related processes: surface heat radiation absorption in atmospheric air, heat storage within soil and heat re-transfer from soil to atmospheric air. The latter two soil processes, although were well-documented in meteorological studies [9,11,18-19], have not been acknowledged to play direct roles in fueling greenhouse effect.

**$CO_2$ concentration-mediated greenhouse effect in atmosphere and soil**

The increased atmospheric air temperatures with $CO_2$ concentration (ranging from 300 ppm to 7500 ppm) at daytime with higher radiation were understandable. Unexpectedly, the magnitude of temperature increase of atmospheric air in mesocosms with 16900 ppm $CO_2$ declined significantly compared to that with 7500 ppm $CO_2$ at daytime with higher radiation. In addition, the temperatures of atmospheric air in mesocosms with substantially higher $CO_2$ concentration (ranging from 3200 ppm to 16900 ppm) were lower than that with the lower $CO_2$ concentration (480 ppm) at early morning and/or nighttime with lower heat radiation. These results emphasized that the molecules of $CO_2$ not only absorb the infrared radiation but also re-emit it to the surrounding space (20). Thus an increase of $CO_2$ concentration in atmospheric air may result in either an increase or decrease of the air temperature in the atmosphere, depending on the balance of heat gain and loss. In other words, $CO_2$ with substantially higher concentration may enhance the net heat loss to colder surrounding interfaces when the heat absorption capacity of $CO_2$ was saturated or heat input was much limited.

As in atmosphere, both a positive and a negative effect of soil $CO_2$ concentration on air temperature in soil were observed. Furthermore, the air temperatures in soil were non-linearly affected by soil $CO_2$ concentration. On one hand, at daytime and early nighttime with higher temperature in soil air, soil air temperature with $CO_2$ concentration of 7500 ppm was higher than that with lower $CO_2$ concentrations. This suggested that an increased $CO_2$ concentration to a limited extent would enhance heat trapping in soil air. On the other hand, the significant decrease of soil air temperature in mesocosms with $CO_2$ concentration of 16900 ppm indicated that soil with substantially higher $CO_2$ concentration may cool the soil probably by transferring more heat to surrounding space during colder periods when the temperature difference between soil and surface atmospheric air became larger. The realistic significance of these findings was greater than those in the atmosphere because $CO_2$ concentration in soil air was often in the range of 1,000 ppm – 20,000 ppm [21-23]. Hence, the variation of soil $CO_2$ concentration may regulate the balance of heat gain and loss in soil which determines the contribution of soil to surface warming of the earth.

The using of polypropylene container has induced some uncertainties in measuring the effects of $CO_2$ concentration on temperature. The temperatures in the simulated air were lower than those in both the real atmosphere and the bulk soil during colder periods such as late night and early morning. These results implied that using of polypropylene container may cause underestimation of the effects of $CO_2$ concentration on air temperature in both the atmosphere and soil when the total heat input into the system was limited. However, this simple mesocosm approach effectively illustrated that the changes of $CO_2$ concentration both in atmosphere and soil may alter the heat balance within respective system.

**Changes in soil and soil $CO_2$ may cause changes in global warming**

Besides the indirect influence of soil on global warming via regulating atmospheric $CO_2$ [24], this study indicates that soil and soil $CO_2$ could contribute directly to the greenhouse effect. In previous studies, soil warming has been recognized as one of consequences or reflections of the atmospheric $CO_2$-induced greenhouse effect [16,19]. Both the average temperature of 10 cm soils and 100 cm soils were increased by 0.31 ℃ decade$^{-1}$ during 1967 - 2002 in the United States [25-26]. However, our results indicated that soil warming, together with the role of atmospheric $CO_2$ in heat radiation trapping, may act as a direct driver of greenhouse effect. If so, the observed increasing soil warming may imply that the warming effect of soils on surface air would be gradually increased. The observed closely positive correlation between surface atmospheric air temperature and soil temperature in the Eurasian continent [18] may also partly reflect such warming effect of soils on surface atmospheric air. In addition, the contribution of soil to greenhouse effect may be manifested in the soil-mediated changes in hydrothermal condition at small spatial scale. For instance, the warming effect of soil on land surface could cause significant difference of microclimate. We have observed that at the foot of a hill, a cement floor was icy but no ice was formed on the adjacent soils. These phenomena implied that, as we postulated in the conceptual model (Fig. 3A), the heat flow from soils which were obstructed by solid surfaces such as cement floor could profoundly change the temperature of land surface.

Given that soil heterogeneity was far greater than that of $CO_2$ distribution in atmosphere, soil heterogeneity-mediated natural greenhouse effect may partly explain the variation of local weather (e.g., large diurnal temperature difference in arid regions) across the earth. In considering that 42% - 68% of the terrestrial land surface had been disturbed during the period 1700 – 2000 [27], we postulated that the anthropogenic changes in soils may be the other important drivers of the globe warming.

Overall, we considered that soil and soil $CO_2$, together with atmospheric $CO_2$, play indispensable roles in fueling the greenhouse effect on the earth. We proposed that anthropogenic changes in soils should be focused in understanding drivers of the globe warming. Further studies are needed to unravel how the human-induced land use changes (e.g., deforestation, farming and cement floor expanding) affect the heat loop among the soil solid phase, water and soil air, and, thus regulate the greenhouse effect.

**METHODS**

**Experimental design**

Eight subplots were chosen at a campus farmland in Henan University (114°18′ E, 34°48′ N), Kaifeng, China, and the Fluvo-aquic soils from each subplot were removed to form a rectanglular pit (length 100 cm × width 80 cm × height 27 cm), with 50 cm distance between any two adjacent subplots. To reduce the potential energy exchange, each of the pits was covered by a polyethylene woven sheet with lining. Then five transparent polypropylene containers (16.5 cm in height, 10.5 cm in diameter) with different levels of $CO_2$ concentration were put on the plastic sheet in a straight line with 3 cm distance between any two adjacent containers. Each of such subplot was regarded as a block; each of such container was regarded as a mesocosm. Half of the eight subplots were randomly chosen and then filled back with the soils, with approximately 10 cm soil covered over the top of the polypropylene containers. Thus, 20 containers were left in the atmosphere of the pits and other 20 containers were buried in soils.

**$CO_2$ concentration manipulation**

The LI840A (Licor, USA) equipped with gas pump (0.5 L min$^{-1}$) was used to manipulate $CO_2$ concentration in the containers at normal atmospheric pressure. In brief, the Helium-oxygen mixture (with 21% $O_2$) was used to replace air in each container at the beginning, and when the $CO_2$ concentration in the container was less than 100 ppm, the pure $CO_2$ was used to prepare $CO_2$ with different concentrations. The five manipulated levels of $CO_2$ concentration were 309 ± 13 ppm (L300), 486 ± 38 ppm (L480), 3203 ± 257 ppm (L3200), 7576 ± 676 ppm (L7500) and 16913 ± 551 ppm (L16900), respectively.

**Temperature monitoring**

The mean air temperature in each container was recorded every ten minutes by temperature sensor of iButton DS1922L (DALLAS, USA) which was hung at the center of the container. Air temperature changes around six days (daytime: 06:01 – 17:51, nighttime: 18:01 – 05:51) were continuously monitored. To assess the bias of temperature measurement derived from the using of polypropylene container, the atmospheric air temperature outside the container at the center of each block were also monitored. To examine the temperature difference between the bulk soil and the simulated soil air, the soil temperature outside the container at the center of each block were also monitored. All the temperature sensors of iButton DS1922L were hung at the same height.

**Data analysis**

According to the pattern of changes in air temperature in atmosphere and soils, data were separated into four different periods for a given day. In brief, periods of early morning with lower heat radiation (Atom-P1), daytime with higher and increasing heat radiation (Atom-P2), daytime with higher and decreasing heat radiation (Atom-P3), and nighttime with lower heat radiation (Atom-P4) were included for the changes of simulated $AT_{atm}$; periods of morning and early afternoon with lower soil air temperature (Soil-P1), daytime and early nighttime with higher and increasing soil air temperature (Soil-P2), daytime and early nighttime with higher and decreasing soil air temperature (Soil-P3), and nighttime with lower soil air temperature (Soil-P4) were included for the changes of simulated $AT_{soil}$.

The repeated measure ANOVA was then performed to explore $CO_2$ concentration effects on the air temperature in mesocosms that were either standing in the atmosphere or covered by 10 cm layer of soils at each period for each of the six experimental days. To assess the biases of temperature measurement derived from the using of polypropylene containers, the temporal change patterns of air temperatures in atmosphere and within containers with similar $CO_2$ concentrations as in surface atmosphere (L300 and L480) were compared; to show the potential heat exchange between soil air and soil particles, the temperature change patterns in the bulk soil and the simulated soil air with five levels of $CO_2$ concentrations were also compared. All statistical analyses were performed with SPSS 19.0 (IBM).

**SUPPLEMENTARY DATA**

Supplementary data are available.

**FUNDING**

This work was supported by the Natural Science Foundation of China (41877054, 31570516), the Zhongyuan Scholar Program (182101510005) and the CAS/SAFEA International Partnership Program for Creative Research Teams.

## AUTHOR CONTRIBUTIONS

W.Z., C.Y., and S.F. initiated the collaborative study and designed the experiment; C.Y., W.Z., Z.S., S.L. and S.L.L. conducted the lab and field work; W.Z. and C.Y. performed data analysis; W.Z., C.Y., Y.S. and S.F. discussed the data and prepared the manuscript.

## ACKNOWLEDGMENTS

We thank G. Li, J.J. Yu and X.X. Chen for help in field experiment, and Drs Xiaoming Zou, Ming Xu, Youming Chen for helpful discussion in manuscript preparation.

## COMPETING INTERESTS

The authors declare no competing interests.

## REFERENCES

1. Lacis AA, Schmidt GA and Rind D *et al*. Atmospheric $CO_2$: principal control knob governing earth's temperature. *Science* 2010; **330**: 356-359.
2. Feldman DR, Collins WD and Gero PJ *et al*. Observational determination of surface radiative forcing by $CO_2$ from 2000 to 2010. *Nature* 2015; **519**: 339-343.
3. Anderson TR, Hawkins E and Jones PD. $CO_2$, the greenhouse effect and global warming: from the pioneering work of Arrhenius and Callendar to today's earth system models. *Endeavour* 2016; **40**: 178-187.

4. Bouwman AF. The role of soils and land use in the greenhouse effect. *Neth J Agr Sci* 1989; **37**: 13-19.

5. Jenkinson DS, Adams DE and Wild A. Model estimates of $CO_2$ emissions from soil in response to global warming. *Nature* 1991; **351**: 304-306.

6. Lal R. Global potential of soil carbon sequestration to mitigate the greenhouse effect. *Crit Rev Plant Sci* 2003; **22**: 151-184.

7. Smith KA, Ball T and Conen F *et al*. Exchange of greenhouse gases between soil and atmosphere: interactions of soil physical factors and biological processes. *Eur J Soil Sci* 2003; **54**: 779-791.

8. Oertel C, Matschullat J and Zurba K *et al*. Greenhouse gas emissions from soils—A review. *Geochemistry* 2016; **76**: 327-352.

9. Di XY. Research of soil thermal properties and it's effects on surface energy balance in Tibet Plateau, thesis, Lanzou University 2009.

10. Zhang H, Wang EL and Zhou DW *et al*. Rising soil temperature in China and its potential ecological impact. *Sci Rep* 2016; **6**: 35530.

11. Tang MC, Sun SH and Zhong Q *et al*. The energy variation of the underlying surface and the changes of the weather and climate. *Plat Meteorol* 1982; **1**: 24-34. (In Chinese with English abstract)

12. Wang XL, Fu SL and Li JX *et al*. Forest soil profile inversion and mixing change the vertical stratification of soil $CO_2$ concentration without altering soil surface $CO_2$ Flux. *Forests* 2019; **10**: 192.

13. Zhao JH, Zhang Q and Wang S *et al*. Effect of soil heat slow transmission process on surface energy balance in semi-arid area. *Chin J Soil Sci* 2013; **44**: 1321-1331. (In Chinese with English abstract)

14. Niu GY, Sun SF and Hong ZX. Numerical simulation on water and heat transport in the desert soil and atmospheric boundary layer. *Acta Meteorol Sin* 1997; **55**: 398-407. (In Chinese with English abstract)

15. Alves M and Soares J. Diurnal variation of soil heat flux at an Antarctic local area during warmer months. *Appl Environ Soil Sci* 2016; **2016**: 1769203.

16. Zhu F, Cuo L and Zhang Y *et al*. Spatiotemporal variations of annual shallow soil temperature on the Tibetan Plateau during 1983–2013. *Clim Dynam* 2018; **51**: 2209-2227.

17. Gebbie G and Huybers P. The Little Ice Age and 20th-century deep Pacific cooling. *Science* 2019; **363**: 70-74.

18. Hu Q and Feng S. How have soil temperatures been affected by the surface temperature and precipitation in the Eurasian continent? *Geophys Res Lett* 2005; **32**: L14711.

19. García-Suárez AM and Butler CJ. Soil temperatures at Armagh Observatory, Northern Ireland, from 1904 to 2002. *Int J Climatol* 2006; **26**: 1075-1089.

20. Shia R. Mechanism of radiative forcing of greenhouse gas and its implication to the global warming. American Geophysical Union Agu Fall Meeting 2010.

21. Hirano T, Kim H and Tanaka Y. Long-term half-hourly measurement of soil $CO_2$ concentration and soil respiration in a temperate deciduous forest. *J Geophys Res Atmos* 2003; **108**: 4631.

22. Wang C, Huang QB and Yang ZJ *et al*. Analysis of vertical profiles of soil $CO_2$ efflux in Chinese fir plantation. *Acta Ecol Sin* 2011; **31**: 5711-5719. (In Chinese with English abstract)

23. Sheng H, Luo S and Zhou P *et al*. Dynamic observation, simulation and application of soil $CO_2$ concentration: a review. *Chin J Appl Ecol* 2012; **23**: 2916-2922. (In Chinese with English abstract)

24. Zhou GY, Xu S and Ciais P *et al*. Climate and litter C/N ratio constrain soil organic carbon accumulation. *Natl Sci Rev* 2019; **0**: 1-12.

25. Hu Q and Feng S. A daily soil temperature dataset and soil temperature climatology of the contiguous United States. *J Appl Meteorol* 2003; **42**: 1139-1156.

26. Hicks Pries CE, Castanha C and Porras R *et al*. Response to Comment on "The whole-soil carbon flux in response to warming". *Science* 2018; **359**: eaao0457.

27. Hurtt GC, Frolking SE and Fearon MG *et al*. The underpinnings of land-use history: three centuries of global gridded land-use transitions, wood-harvest activity, and resulting secondary lands. *Global Change Biol* 2006; **12**: 1208-1229.

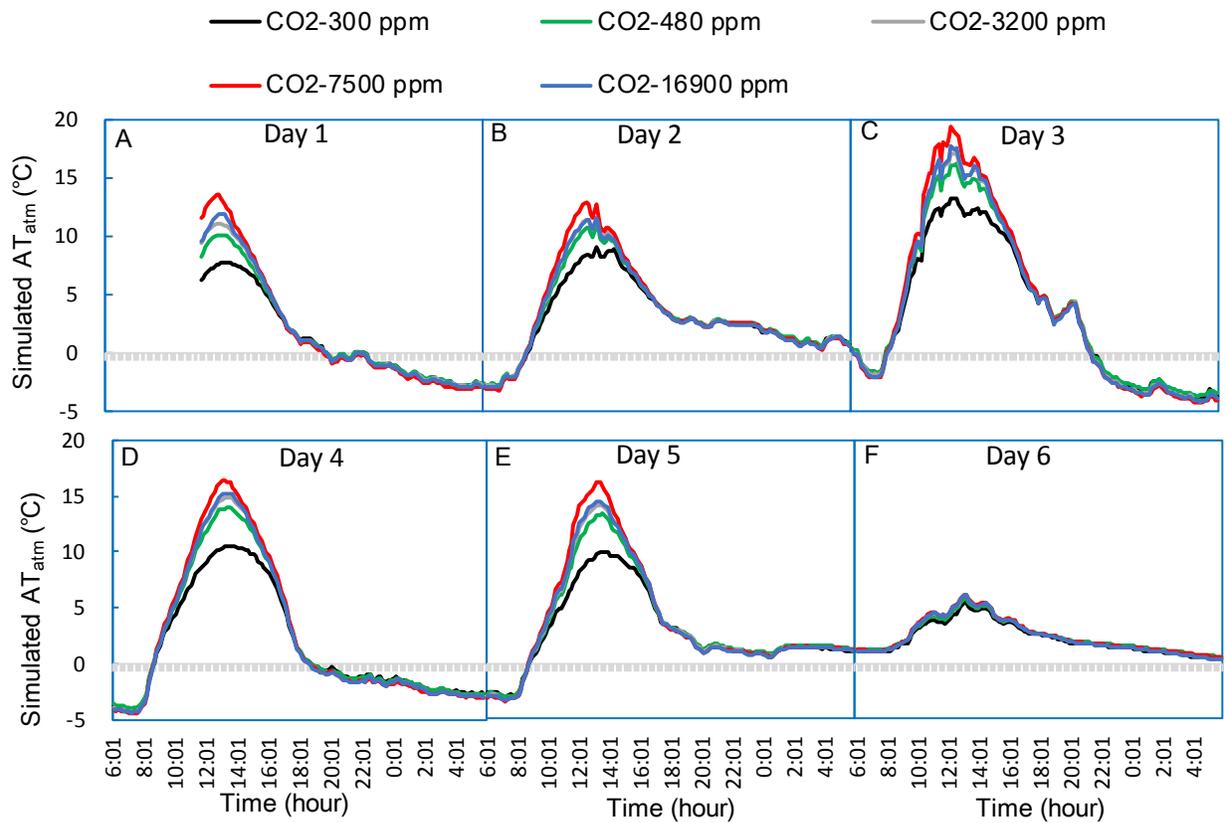

**Fig. 1**. The effects of $CO_2$ concentration on the temperatures in simulated atmospheric air during the six experimental days. $AT_{atm}$: air temperature in atmosphere.

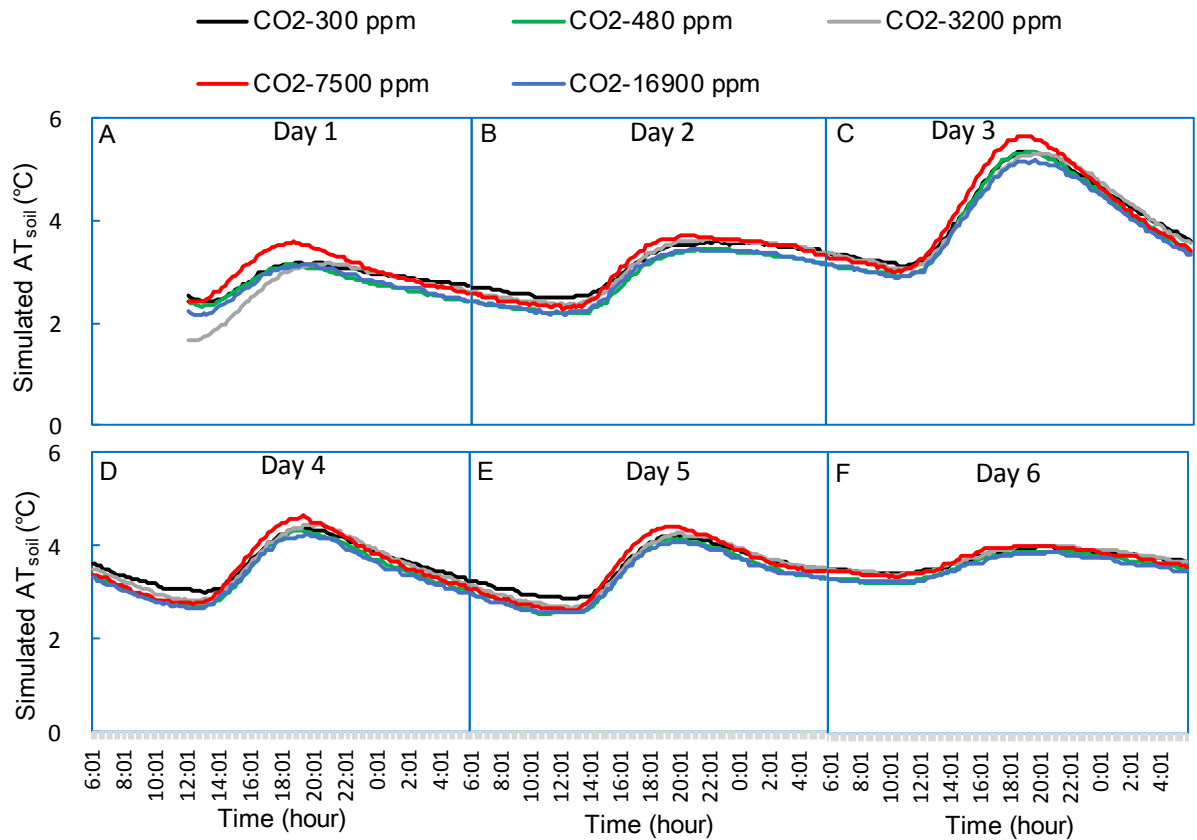

**Fig. 2**. The effects of $CO_2$ concentration on the temperatures in simulated soil air (covered by 10 cm surface soil) during the six experimental days. $AT_{soil}$: air temperature in soil.

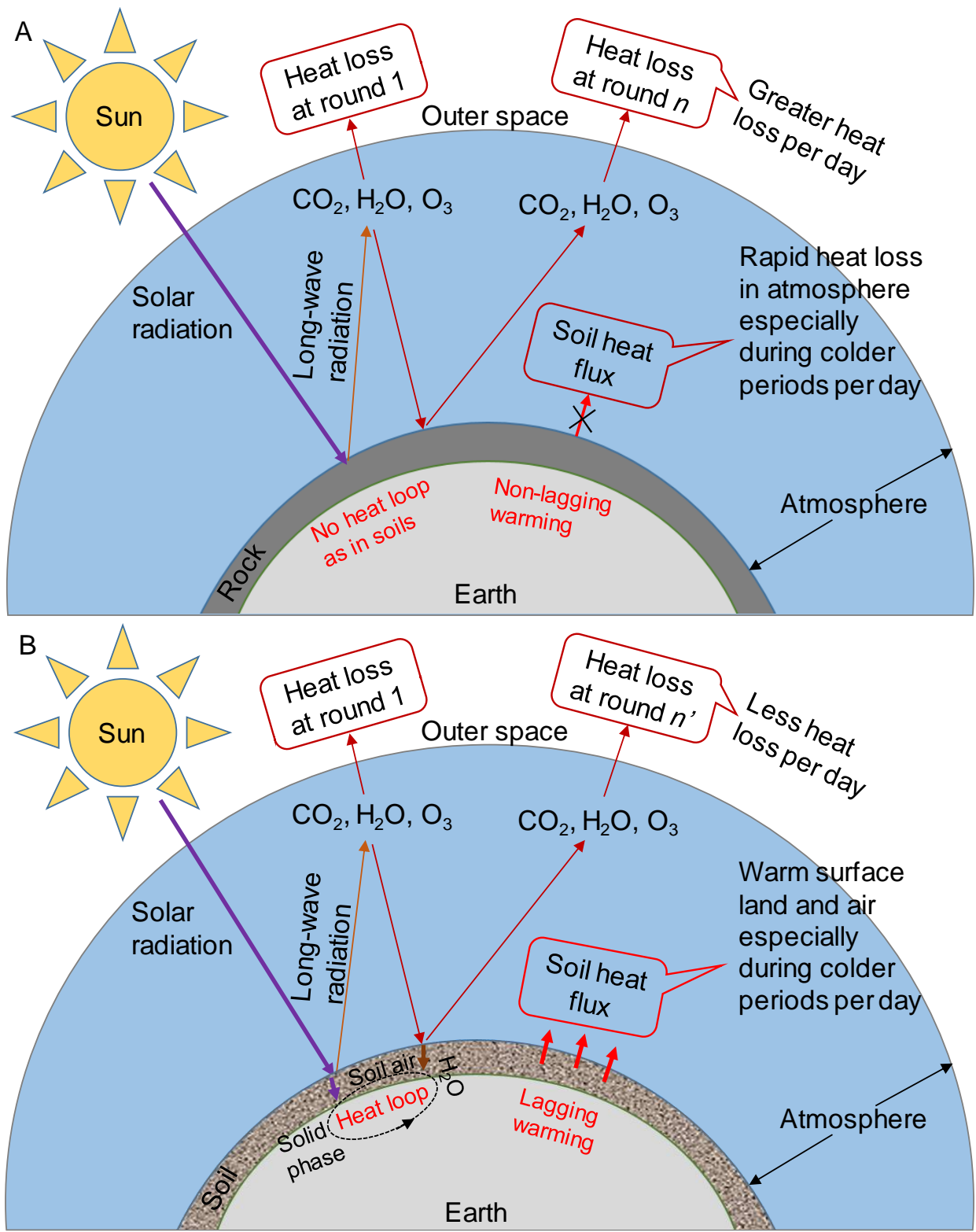

Fig. 3. A conceptual model showing how soil contribute to greenhouse effect. Panel (**A**) assuming only the bare rock remained on the surface of the terrestrial lands. Panel (**B**) showing

natural lands with soils. "×" means the heat flux disappeared. The "heat loop" represents the potential heat transfer among the three major components of soil, i.e., soil solid phase, soil air and soil water. The short white-yellow arrows refer to heat flow being transferred into soil or out to the surface atmospheric air. The $n$ and $n'$ refers to the total number of $CO_2$-mediated heat absorption-release round per day in terrestrial lands covered with bare rock and soils, respectively; at daily scale, $n' < n$ due to the slower heat transmission process within soil compared to that in atmosphere.

# Supplementary Materials for

## Soil and soil $CO_2$ magnify greenhouse effect

Weixin Zhang[1†], Chengde Yu[2†], Zhifeng Shen[1], Shu Liu[1], Suli Li[1], Yuanhu Shao[1], Shenglei Fu[1*].

Correspondence to: sfu@scbg.ac.cn

**This file includes:**

Figs. S1 to S3

Tables S1 to S2

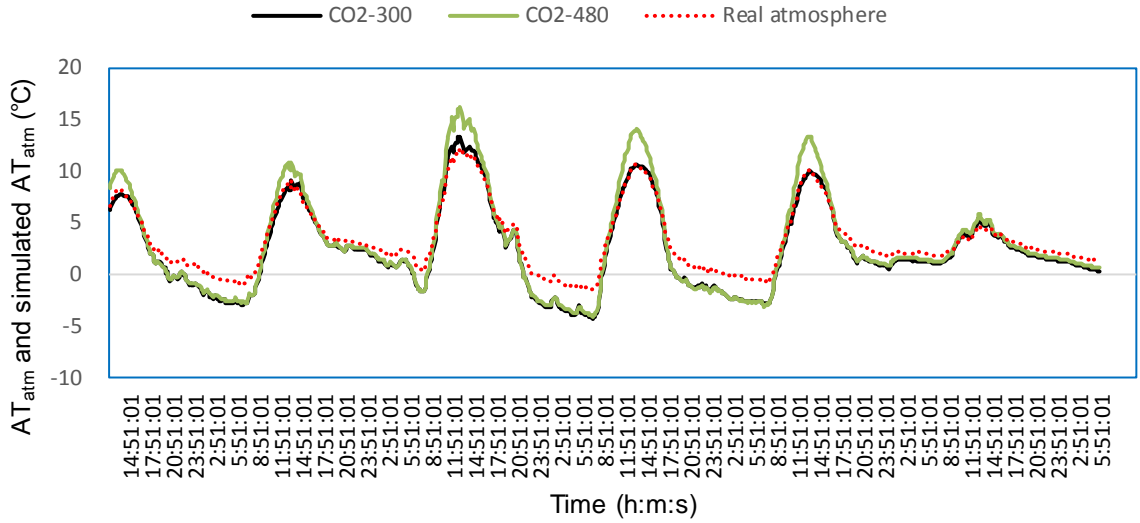

**Fig. S1.** The comparisons of change patterns of temperatures in the real atmospheric air ($AT_{atm}$) at block center and in the simulated atmospheric air with $CO_2$ concentration of 300 ppm and 480 ppm during the six experimental days.

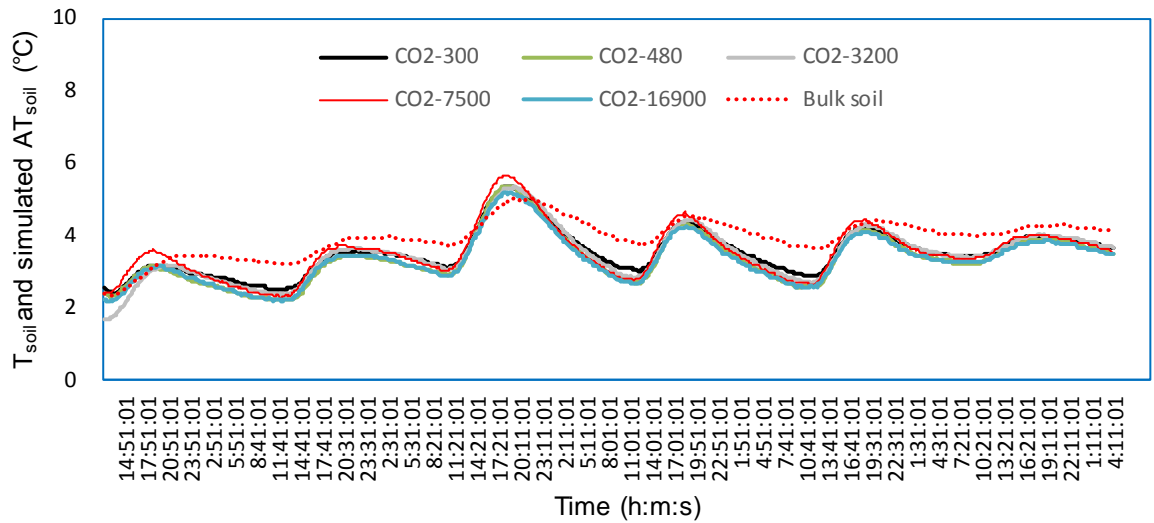

**Fig. S2.** The comparisons of change pattern of temperatures in bulk soil ($T_{soil}$) at block center and in the simulated soil air ($AT_{soil}$) with five levels of $CO_2$ concentration (300 ppm – 16900 ppm) during the six experimental days.

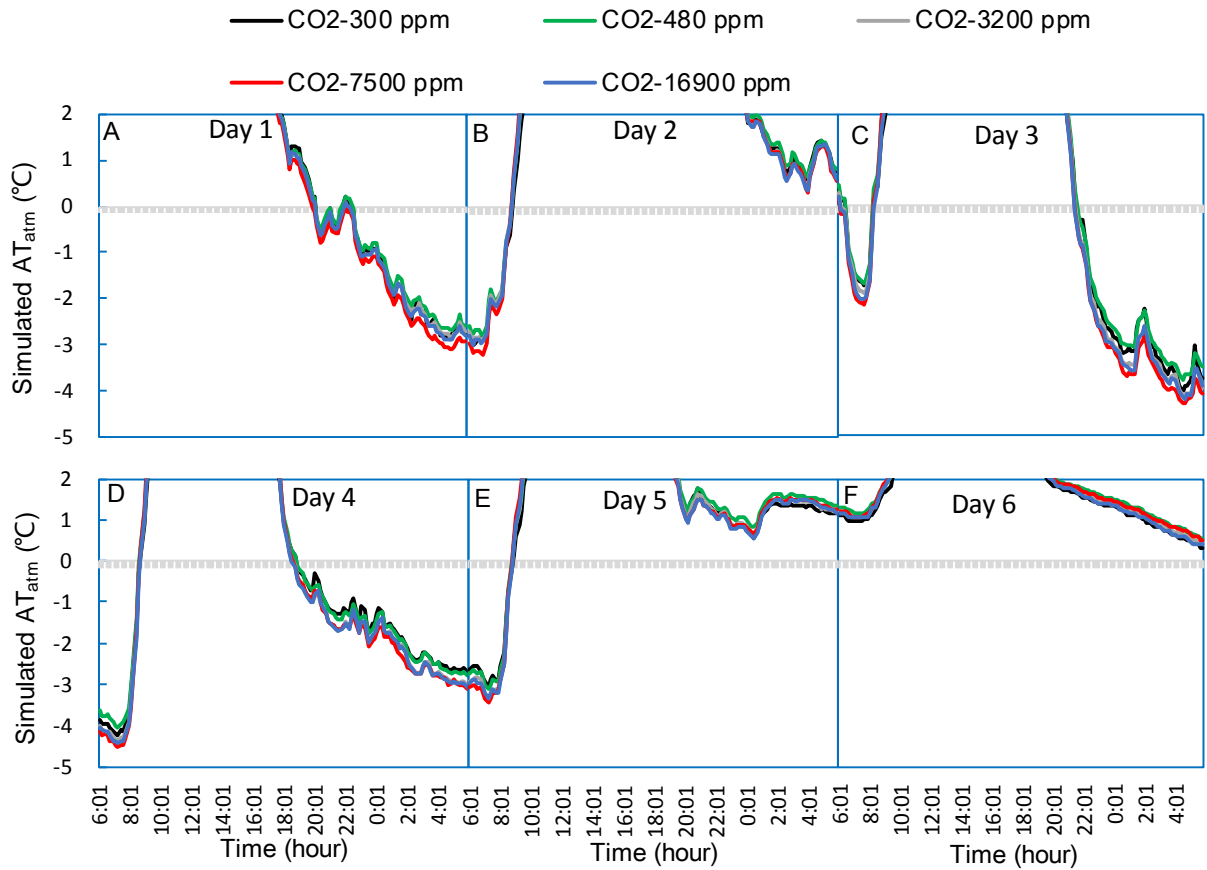

**Fig. S3.** The negative effect of increased $CO_2$ concentration on temperature in the simulated atmospheric air at nighttime and early morning during which the environmental temperature was lower than 2℃.

1　**Table S1.** The repeated measure ANOVA results of $CO_2$ concentration effects on the temperature in the simulated atmospheric air
2　during the six experimental days. NA: data not available; †noted that the air temperature in mesocosms with treatment of L480 was
3　likely to be higher than that of L7500 and L16900 ($P = 0.051$ and $P = 0.079$, respectively); ‡noted that the air temperature in
4　mesocosms with treatment of L480 was likely to be higher than that of L300 ($P = 0.072$); ¶ noted that the air temperature in
5　mesocosms with treatment of L300 was likely to be lower than that of L3200 and L7500 ($P = 0.079$ and $P = 0.079$, respectively);
6　§noted that the air temperature in mesocosms with treatment of L300 was likely to be lower than that of L3200 and L7500 ($P = 0.058$
7　and $P = 0.074$, respectively); ‖noted that the air temperature in mesocosms with treatment of L300 was likely to be lower than that of
8　L480 ($P = 0.051$). The upward and downward arrows indicated that some of the treatments with higher $CO_2$ concentration may cause
9　increased and decreased air temperature in the mesocosms, respectively; the arrow with dashed line indicated that the $CO_2$
10　concentration effect on air temperature in mesocosms was not significant but deserved to be noticed ($0.05 < P < 0.1$). The difference
11　letter indicated significant effect of $CO_2$ concentration on air temperature ($P < 0.05$).

| Days | Periods | Time (h: m) | Air temperature in mesocosms (℃, Mean ± SE) | | | | | $F$ and $P$ value |
| --- | --- | --- | --- | --- | --- | --- | --- | --- |
| | | | Levels of $CO_2$ concentration in mesocosms (ppm) | | | | | |
| | | | L300 | L480 | L3200 | L7500 | L16900 | |
| Day 1 | Atom-P1: Early morning with lower heat radiation | NA | NA | NA | NA | NA | NA | NA |
|  | Atom-P2: Daytime with higher and increasing heat radiation | 12:01-13:11 | 7.06 ± 0.18c | 9.44 ± 0.24b | 10.49 ± 0.22b | 12.77 ± 0.27a | 10.87 ± 0.31b | $F_{4,15} = 12.62$; $P < 0.001$; ↑↓ |
|  | Atom-P3: Daytime with higher and decreasing heat radiation | 13:21-16:41 | 6.56 ± 0.26c | 7.73 ± 0.41b | 8.27 ± 0.47ab | 8.97 ± 0.59a | 8.51 ± 0.51ab | $F_{4,15} = 11.83$; $P < 0.001$; ↑ |
|  | Atom-P4: Nighttime with lower heat radiation | 0:01-5:51 | -2.19 ± 0.09ab | -2.05 ± 0.09a | -2.19 ± 0.09ab | -2.43 ± 0.10b | -2.24 ± 0.09ab | $F_{4,15} = 1.86$; $P = 0.170$; ↓ |
| Day 2 | Atom-P1: Early morning with lower heat radiation | 6:01-8:01 | -2.57 ± 0.11ab | -2.41 ± 0.10a | -2.53 ± 0.11ab | -2.78 ± 0.12b | -2.60 ± 0.11ab | $F_{4,15} = 1.80$; $P = 0.182$; ↓ |
|  | Atom-P2: Daytime with higher and increasing heat radiation | 9:31-13:21 | 6.03 ± 0.44c | 7.58 ± 0.57b | 8.14 ± 0.60b | 9.16 ± 0.68a | 8.17 ± 0.60b | $F_{4,15} = 14.92$; $P < 0.001$; ↑↓ |
|  | Atom-P3: Daytime with higher and decreasing heat radiation | 13:31-16:31 | 7.40 ± 0.28d | 8.10 ± 0.37c | 8.44 ± 0.40ab | 8.67 ± 0.44a | 8.23 ± 0.39bc | $F_{4,15} = 24.78$; $P < 0.001$; ↑↓ |
|  | Atom-P4: Nighttime with lower heat radiation | 0:01-5:51 | 1.22 ± 0.07ab | 1.31 ± 0.07a | 1.15 ± 0.07b | 1.14 ± 0.08b | 1.12 ± 0.08b | $F_{4,15} = 2.54$; $P = 0.083$; ↓ |
| Day 3 | Atom-P1: Early morning with lower heat radiation | 6:01-8:21 | -0.91 ± 0.20a | -0.84 ± 0.21a | -1.04 ± 0.21ab | -1.22 ± 0.23b | -1.17 ± 0.22b | $F_{4,15} = 4.40$; $P = 0.015$; ↓ |
| | Atom-P2: Daytime with higher and increasing heat radiation | 9:21-12:31 | 9.66 ± 0.69c | 11.32 ± 0.86b | 12.00 ± 0.93b | 13.30 ± 1.05a | 12.02 ± 0.96b | $F_{4,15} = 11.63$; $P < 0.001$; ↑↓ |
| | Atom-P3: Daytime with higher and decreasing heat radiation | 12:41-15:21 | 12.07 ± 0.15c | 14.33 ± 0.26b | 15.02 ± 0.30ab | 15.95 ± 0.42a | 15.15 ± 0.34ab | $F_{4,15} = 9.44$; $P = 0.001$; ↑ |
| | Atom-P4: Nighttime with lower heat radiation | 0:00-5:51 | -3.26 ± 0.08ab | -3.17 ± 0.07a | -3.48 ± 0.07ab | -3.70 ± 0.07b | -3.53 ± 0.07ab | $F_{4,15} = 1.77$; $P = 0.188$; ↓ |
| Day 4 | Atom-P1: Early morning with lower heat radiation | 6:01-8:31 | -3.50 ± 0.28a | -3.31 ± 0.28a | -3.66 ± 0.28a | -3.77 ± 0.30a | -3.72 ± 0.27a | $F_{4,15} = 1.49$; $P = 0.254$†; ↑⋮↓ |
| | Atom-P2: Daytime with higher and increasing heat radiation | 9:11-13:11 | 7.11 ± 0.53c | 9.01 ± 0.75b | 9.56 ± 0.81ab | 10.40 ± 0.90a | 9.62 ± 0.83ab | $F_{4,15} = 7.68$; $P = 0.001$; ↑ |
| | Atom-P3: Daytime with higher and decreasing heat radiation | 13:21-17:01 | 8.90 ± 0.37c | 10.74 ± 0.56ab | 11.25 ± 0.62a | 11.79 ± 0.68a | 11.25 ± 0.66a | $F_{4,15} = 9.30$; $P = 0.001$; ↑ |
| | Atom-P4: Nighttime with lower heat radiation | 0:01-5:51 | -2.18 ± 0.08a | -2.23 ± 0.08a | -2.44 ± 0.08a | -2.52 ± 0.07a | -2.46 ± 0.08a | $F_{4,15} = 0.367$; $P = 0.828$ |
| Day 5 | Atom-P1: Early morning with lower heat radiation | 6:01-8:41 | -2.82 ± 0.23a | -2.41 ± 0.21a | -2.60 ± 0.22a | -2.69 ± 0.23a | -2.65 ± 0.21a | $F_{4,15} = 0.327$; $P = 0.856$ |

|  | Atom-P2: Daytime with higher and increasing heat radiation | 9:51-13:11 | 7.96 ± 0.65b | 8.74 ± 0.73b | 9.43 ± 0.80ab | 10.71 ± 0.93a | 9.70 ± 0.82ab | $F_{4,15}$ = 2.80; $P$ = 0.064; ↑ |
|  | Atom-P3: Daytime with higher and decreasing heat radiation | 13:21-16:21 | 10.08 ± 0.35b | 10.76 ± 0.45ab | 11.37 ± 0.49ab | 11.98 ± 0.63a | 11.36 ± 0.51ab | $F_{4,15}$ = 2.26; $P$ = 0.112; ↑ |
|  | Atom-P4: Nighttime with lower heart radiation | 0:01-5:51 | 1.25 ± 0.04b | 1.42 ± 0.04a | 1.29 ± 0.04ab | 1.31 ± 0.04ab | 1.26 ± 0.05b | $F_{4,15}$ = 1.81; $P$ =0.181, ↑↓ |
| Day 6 | Atom-P1: Early morning with lower heat radiation | 6:01-7:41 | 1.10 ± 0.02a | 1.25 ± 0.02a | 1.13 ± 0.01a | 1.15 ± 0.02a | 1.12 ± 0.01a | $F_{4,15}$ = 1.13; $P$ =0.379‡, ↑⋮ |
|  | Atom-P2: Daytime with higher and increasing heat radiation | 11:21-13:01 | 4.60 ± 0.19a | 4.62 ± 0.19a | 4.92 ± 0.20a | 4.92 ± 0.20a | 4.89 ± 0.20a | $F_{4,15}$ = 1.90; $P$ =0.162¶, ↑⋮ |
|  | Atom-P3: Daytime with higher and decreasing heat radiation | 13:11-14:51 | 5.14 ± 0.09a | 5.17 ± 0.09a | 5.42 ± 0.11a | 5.41 ± 0.11a | 5.38 ± 0.11a | $F_{4,15}$ = 1.96; $P$ =0.152§, ↑⋮ |
|  | Atom-P4: Nighttime with lower heat radiation | 0:01-5:51 | 0.92 ± 0.05ab | 1.05 ± 0.05a | 0.93 ± 0.05ab | 1.01 ± 0.05a | 0.87 ± 0.05b | $F_{4,15}$ = 1.81; $P$ =0.181‖, ↑↓⋮ |

Table S2. The repeated measure ANOVA results of $CO_2$ concentration effects on the air temperature of the simulated soil air in mesocosms covered by 10 cm layer of soil during the six experimental days. †noted that the air temperature in mesocosms with treatment of L7500 was likely to be higher than that of L300 ($P = 0.06$); ‡noted that the air temperature in mesocosms with treatment of L7500 was likely to be higher than that of L3200 ($P = 0.073$). NA: data not available. The upward and downward arrows indicated that some of the treatments with higher $CO_2$ concentration may cause increased and decreased air temperature in the mesocosms, respectively. The difference letter indicated significant effect of $CO_2$ concentration on air temperature ($P < 0.05$).

| Days | Periods | Time (h: m) | Air temperature in mesocosms (℃, Mean ± SE) | | | | | F and P value |
|---|---|---|---|---|---|---|---|---|
| | | | Levels of $CO_2$ concentration in mesocosms (ppm) | | | | | |
| | | | L300 | L480 | L3200 | L7500 | L16900 | |
| Day 1 | Soil-P1: Morning and early afternoon with lower soil air temperature | NA | NA | NA | NA | NA | NA | NA |
| | Soil-P2: Daytime and early nighttime with higher and increasing soil air temperature | 14:01-18:41 | 2.86 ± 0.06ab | 2.85 ± 0.11ab | 2.54 ± 0.08b | 3.22 ± 0.06a | 2.78 ± 0.04ab | $F_{4,15} = 1.56$; $P = 0.236$; ↑ |
| | Soil-P3: Daytime and early nighttime with higher and decreasing soil air temperature | 18:51-0:31 | 3.09 ± 0.02ab | 2.93 ± 0.04b | 3.09 ± 0.05ab | 3.27 ± 0.03a | 2.98 ± 0.02b | $F_{4,15} = 2.30$; $P = 0.106$; ↑↓ |
| | Soil-P4: Nighttime with lower soil air temperature | 0:41-5:51 | 2.83 ± 0.03a | 2.56 ± 0.03a | 2.77 ± 0.05a | 2.75 ± 0.05a | 2.59 ± 0.02a | $F_{4,15} = 1.08$; $P = 0.40$ |
| Day 2 | Soil-P1: Morning and early afternoon with lower soil air temperature | 6:01-12:01 | 2.58 ± 0.03a | 2.28 ± 0.04a | 2.46 ± 0.05a | 2.41 ± 0.06a | 2.29 ± 0.03a | $F_{4,15} = 0.89$; $P = 0.493$ |
| | Soil-P2: Daytime and early nighttime with higher and increasing soil air temperature | 15:21-21:01 | 3.29 ± 0.04ab | 3.11 ± 0.02b | 3.30 ± 0.04ab | 3.44 ± 0.04a | 3.16 ± 0.01b | $F_{4,15} = 2.22$; $P = 0.116$; ↑↓ |
| | Soil-P3: Daytime and early nighttime with higher and decreasing soil air temperature | 21:11-1:51 | 3.55 ± 0.05a | 3.40 ± 0.03a | 3.59 ± 0.06a | 3.63 ± 0.04a | 3.40 ± 0.03a | $F_{4,15} = 0.82$; $P = 0.531$ |
| | Soil-P4: Nighttime with lower soil air temperature | 2:01-5:51 | 3.45 ± 0.05a | 3.24 ± 0.04a | 3.47 ± 0.07a | 3.44 ± 0.06a | 3.26 ± 0.04a | $F_{4,15} = 0.70$; $P = 0.604$ |
| Day 3 | Soil-P1: Morning and early afternoon with lower soil air temperature | 6:01-10:31 | 3.26 ± 0.05a | 3.03 ± 0.04a | 3.22 ± 0.07a | 3.16 ± 0.06a | 3.02 ± 0.04a | $F_{4,15} = 0.63$; $P = 0.649$ |
| | Soil-P2: Daytime and early nighttime with higher and increasing soil air temperature | 12:11-18:41 | 4.33 ± 0.02b | 4.28 ± 0.04b | 4.23 ± 0.02b | 4.56 ± 0.02a | 4.18 ± 0.03b | $F_{4,15} = 4.17$; $P = 0.018$; ↑↓ |
| | Soil-P3: Daytime and early nighttime with higher and decreasing soil air temperature | 18:51-22:11 | 5.21 ± 0.03ab | 5.19 ± 0.03b | 5.24 ± 0.05ab | 5.42 ± 0.03a | 5.07 ± 0.02b | $F_{4,15} = 3.04$; $P = 0.051$†; ↑↓ |
| | Soil-P4: Nighttime with lower soil air temperature | 0:01-5:51 | 4.08 ± 0.04a | 3.91 ± 0.04a | 4.10 ± 0.07a | 3.96 ± 0.06a | 3.85 ± 0.05a | $F_{4,15} = 0.47$; $P = 0.758$ |
| Day 4 | Soil-P1: Morning and early afternoon with lower soil air temperature | 6:01-12:51 | 3.24 ± 0.04a | 2.94 ± 0.05a | 3.08 ± 0.06a | 2.98 ± 0.06a | 2.91 ± 0.06a | $F_{4,15} = 0.71$; $P = 0.599$ |
| | Soil-P2: Daytime and early nighttime with higher and increasing soil air temperature | 14:41-19:21 | 3.92 ± 0.03ab | 3.82 ± 0.02b | 3.89 ± 0.04ab | 4.11 ± 0.04a | 3.78 ± 0.01b | $F_{4,15} = 2.50$; $P = 0.087$‡; ↑↓ |
| | Soil-P3: Daytime and early nighttime with higher and decreasing soil air temperature | 19:31-22:31 | 2.44 ± 0.63a | 2.49 ± 0.60a | 2.47 ± 0.65a | 2.55 ± 0.66a | 2.22 ± 0.65a | $F_{4,15} = 0.02$; $P = 0.999$ |
| | Soil-P4: Nighttime with lower soil air temperature | 0:01-5:51 | 1.49 ± 0.49a | 1.46 ± 0.46a | 1.45 ± 0.48a | 1.52 ± 0.50a | 1.22 ± 0.49a | $F_{4,15} = 0.02$; $P = 0.999$ |
| Day 5 | Soil-P1: Morning and early afternoon with lower soil air temperature | 6:01-13:11 | 3.02 ± 0.03a | 2.71 ± 0.04a | 2.86 ± 0.05a | 2.78 ± 0.06a | 2.70 ± 0.05a | $F_{4,15} = 0.73$; $P = 0.584$ |
| | Soil-P2: Daytime and early nighttime with higher and increasing soil air temperature | 14:41-20:01 | 3.81 ± 0.03ab | 3.71 ± 0.02b | 3.77 ± 0.04ab | 3.99 ± 0.04a | 3.66 ± 0.01b | $F_{4,15} = 2.49$; $P = 0.087$; ↑↓ |
| | Soil-P3: Daytime and early nighttime with higher and decreasing soil air temperature | 20:11-23:31 | 4.09 ± 0.06a | 4.01 ± 0.04a | 4.17 ± 0.07a | 4.21 ± 0.06a | 3.96 ± 0.04a | $F_{4,15} = 0.75$; $P = 0.572$ |
| | Soil-P4: Nighttime with lower soil air temperature | 0:01-5:51 | 3.66 ± 0.05a | 3.49 ± 0.04a | 3.68 ± 0.06a | 3.62 ± 0.06a | 3.48 ± 0.05a | $F_{4,15} = 0.38$; $P = 0.819$ |
| Day 6 | Soil-P1: Morning and early afternoon with lower soil air temperature | 6:01-11:41 | 3.44 ± 0.05a | 3.24 ± 0.04a | 3.43 ± 0.06a | 3.40 ± 0.05a | 3.23 ± 0.05a | $F_{4,15} = 0.49$; $P = 0.741$ |
| | Soil-P2: Daytime and early nighttime with higher and increasing soil air temperature | 11:51-21:01 | 3.77 ± 0.03a | 3.67 ± 0.03a | 3.77 ± 0.03a | 3.81 ± 0.03a | 3.63 ± 0.03a | $F_{4,15} = 0.48$; $P = 0.750$ |
| | Soil-P3: Daytime and early nighttime with higher and decreasing soil air temperature | 21:11-23:51 | 3.92 ± 0.06a | 3.84 ± 0.06a | 3.95 ± 0.07a | 3.93 ± 0.06a | 3.78 ± 0.06a | $F_{4,15} = 0.34$; $P = 0.850$ |
| | Soil-P4: Nighttime with lower soil air temperature | 0:01-5:51 | 3.78 ± 0.04a | 3.64 ± 0.05a | 3.79 ± 0.05a | 3.71 ± 0.05a | 3.59 ± 0.04a | $F_{4,15} = 0.36$; $P = 0.830$ |